\documentclass[useAMS,usenatbib,usegraphicx]{mn2e}

\title[Age determination of the HR\,8799 planetary system using asteroseismology]{Age determination of the HR\,8799 planetary system using asteroseismology}
\author[A. Moya et al.]{A. Moya$^{1}$\thanks{E-mail:
amoya@cab.inta-csic.es}, P. J. Amado$^{2}$, D. Barrado$^{3,1}$,
  A. Garc\'{\i}a Hern\'andez$^{2}$, \and M. Aberasturi$^{1}$,
  B. Montesinos$^{1}$, F. Aceituno$^{2}$\\
$^{1}$Departamento de Astrof\'{\i}sica, Laboratorio de Astrof\'isica
  Estelar y Exoplanetas, LAEX-CAB (INTA-CSIC), PO BOX 78,\\ 
28691 Villanueva de la Ca\~nada, Madrid, Spain\\
$^{2}$Instituo de Astrof\'{\i}sica de Andaluc\'{\i}a - CSIC, Camino Bajo de Huetor 50, 18008, Granada, Spain\\
$^{3}$Calar Alto Observatory, German-Spanish Astronomical Center, C/ Jes\'us
  Durb\'an Rem\'on, 2-2, 04004, Almer\'ia, Spain}
\begin{document}



\maketitle

\label{firstpage}

\begin{abstract}
Discovery of the first planetary system by direct imaging around
HR\,8799 has made the age determination of the host star a very
important task. This determination is the key to derive accurate
masses of the planets and to study the dynamical stability of the
system. The age of this star has been estimated using different
procedures. In this work we show that some of these procedures have
problems and large uncertainties, and the real age of this star is
still unknown, needing more observational constraints. Therefore, we
have developed a comprehensive modeling of HR\,8799, and taking
advantage of its $\gamma$ Doradus-type pulsations, we have estimated
the age of the star using asteroseismology. The accuracy in the age
determination depends on the rotation velocity of the star, and
therefore an accurate value of the inclination angle is required to
solve the problem. Nevertheless, we find that the age estimate for
this star previously published in the literature ([30,160] Myr) is
unlikely, and a more accurate value might be closer to the Gyr.  This
determination has deep implications on the value of the mass of the
objects orbiting HR\,8799. An age around $\approx$1 Gyr implies that
these objects are brown dwarfs.
\end{abstract}


\section{Introduction}

The discovery of the first planetary system by direct imaging around
HR\,8799 \citep{Marois} was an important milestone in the field of
exoplanet research. Up to now, eleven planets have been discovered
with this procedure, and only one possible planetary system. One of
the advantages of this technique is the direct measurement of the
luminosity and projected orbits of the planets, making the physical
characterization of the system and the individual planets possible.

\cite{Marois} used a procedure for estimating the mass of the objects
around HR\,8799 in order to discriminate whether they were real
planets or brown dwarfs (BD). This procedure can be applicable to any
direct imaging detection, and it is based on the comparison of
theoretical evolutionary tracks of BD and giant planets with
observations in an Age - Luminosity diagram. Since luminosity is a
direct observable, this technique is limited by the accuracy of the
age determination and the theoretical models used for this comparison
\citep{reider}.

The A5V spectral type star HR\,8799 (V342 Peg, HD\,218396,
HIP\,114189) has been extensively studied. \cite{schuster} firstly
reported this star as a possible SX Phoenicis object (these stars are
pulsating subdwarf stars with periods larger than one
day). \cite{zerbi} classified it as one of the 12 first $\gamma$
Doradus pulsators known. This pulsating stellar type is composed of
Main Sequence (MS) stars in the lower part of the classical
instability strip \citep{tdcma}, with periods around one day. That
means that their pulsating modes are asymptotic $g$-modes and they are
suitable to study the deep interior of the star. \cite{gray} obtained
an optical spectrum of HR\,8799, and assigned an spectral type of kA5
hF0 mA5 V $\lambda$ Bootis (see that paper for the meaning of this
specific nomenclature), reporting a metallicity of [M/H]$=-0.47$. The
$\lambda$ Bootis nature of the star means that it has solar surface
abundances of light elements, and subsolar abundances of heavy
elements, the internal metallicity of the star being unknown. They
also noted that HR\,8799 may be also a Vega-type star, characterized
by a far IR excess due to a debris disk. Up to now, only three
$\lambda$ Bootis stars have been reported to be $\gamma$ Doradus
pulsators: HD\,218427 \citep{rodriguez06a}, HD\,239276
\citep{rodriguez06b}, and HR\,8799.

\cite{Marois} estimated the age of the star using four different
methods (see Section 2 for details). The conclusion of that work was
that HR\,8799 is a young MS star with an age in the range $[30, 160]$
Myr. \cite{reider} added another element for estimating the age of
this star: the infrared excess ratio, studied by \cite{su2006} and
also used by \cite{chen}. None of these determinations are conclusive,
as we explain in Section 2.

In this work, a comprehensive modelling of HR\,8799 has been done in
order to estimate its age and mass. The $\lambda$ Bootis nature of
this star, and its $\gamma$ Doradus pulsations are the bases of the
determinations presented here. All the technical details are presented
in an accompanying paper \citep[Paper I]{hr8799lambda}, where, a
complementary study on the $\lambda$ Bootis nature of this star is
developed.

\section{Previous age determinations of HR\,8799}

Most of the procedures to determine the age of isolated stars in the
literature, such as activity, surface lithium abundance, or the
position in the HR diagram, are useful for young and for low-mass
stars, however these procedures cannot be accurately applied to an A5
MS star.


\begin{table}
 \centering
 \begin{minipage}{80mm}
  \caption{Age of HR\,8799 in the literature. The methods used
      are: a) Stellar kinematics groups (proper motions), b) HR
      diagram position compared with isochrones, c) HR diagram position
      compared with models, d) IR excess.}
  \begin{tabular}{ccc}
 \hline
Age range & Reference & Method used\\
in Myr & & \\
\hline
$[20,150]$ &  \cite{moor} & a\\
$[30,160]$ &  \cite{Marois} & a, b\\
$[50,1128]$ &  \cite{song} & c\\
30 & \cite{zuck} & a, b\\
30 & \cite{rhee} & a, b\\
$[30,730]$ & \cite{chen} & a, b, d\\
\hline
\end{tabular}
\end{minipage}
\end{table}

\cite{kaye2} studied the chromospheric activity of some $\gamma$
Doradus stars, in particular HR\,8799, using the Ca II H\&K lines,
and found that all the stars studied have very low activity,
HR\,8799 being the least active of all. This conclusion has an impact
in its age determination since activity decreases with age in the
early evolutionary stages of the star \citep{soderblom}. The very low
activity of HR\,8799 suggests that it is a Main Sequence star.

HR\,8799 has a debris disk that has been used for an estimation of its
age. \cite{zuck,moor} studied the age of a number of stars with debris
disks, searching for a relationship between the fractional IR
luminosity $L_{\rm{IR}}/L_{\rm{bol}}$ and the age. HR\,8799 was
present in the samples covered in these works; they estimated an age
of 30 Myr \citep{zuck,rhee} or in the range [20, 150] Myr
\citep{moor}. Note that this age estimation was purely statistical,
and has to be taken it with caution. In addition, the debris disks
have cataclismic origins (i.e. highly chaotic and unpredictable), and
hardly can be related with the age of the star.

\cite{Marois} used four methods for estimating the age of
HR\,8799. The first one was the measurement of its radial velocity and
proper motion (UVW). Comparing with other stars and associations with
similar values of these quantities, they found an age range similar to
that reported in previous works using also proper motions ([30, 160]
Myr). They warned that this method is not always reliable.

The  second method  was the  position of  the star  in the  HR diagram
\citep[for  instance,  see][]{pont}.   Comparing  this  position  with
isochrones of known stellar clusters one can infer upper limits to the
age of  the star. This argument  was used to confirm  the age obtained
with the  statistical UVW  method. Nevertheless, the  $\lambda$ Bootis
nature of this star, and its unknown internal metalicity, makes the HR
diagram position of  the star an inaccurate method  for the estimation
of the star. \cite{song} studied the  age of a set of A-type stars (in
particular  HR\,8799)  using  Str\"omgren photometry  and  isochrones,
taking  into  account  the  stellar  rotation.  They  concluded  that,
regarding the position  in the HR diagram, the  age estimated for this
star is in the range [50,1128] Myr.

The third method was based on the fact that $\lambda$ Bootis stars are
generaly young. This argument is not correct.  \citet{Paunzeniliev02}
found that most of the known $\lambda$ Bootis stars are stars between
the ZAMS (zero-age main sequence) and the TAMS (terminal-age main
sequence), with a mean age of 1 Gyr.

The fourth method was based on the assumption that $\gamma$ Doradus
pulsators are probably young stars. There is no evidence for this
assertion.  $\gamma$ Doradus pulsations are originated from the
blocking of the radiative flux at the base of the outer convective
zone \citep{guzik}. Therefore, it is mainly a thermodinamic effect,
dominated by the temperature of the star and the efficiency of the
convective transport. The existence, or not, of $\gamma$ Doradus
pulsations does not provide information about the age of the star, at
least for a range in ages within the main sequence.

Table 1 shows a summary of the different age determinations
of this star found in the literature.

\section{Asteroseismological estimation of the age of HR\,8799}

In the previous section we indicated that there is no method that can
clearly provide the age of the HR\,8799 planetary system. Of all the
methods mentioned in the previous section, three seems to support a
young age for our star. Of those, we will only consider those based on
proper motions and on the position in the HR diagram, the third being
an statistical method (viz., the luminosity of debris disks) with
little accuracy in age determination. Our first step is to revise the
physical parameters of the star and its position in the HR diagram
with the aim of checking whether the overlap in ages provided by this
method and the proper motions is supported

\begin{table}
 \centering
  \caption{Physical characteristics of HR\,8799}
  \begin{tabular}{rrr}
 \hline
$T_{\rm{eff}}$ (K) & 7430$\pm$75 & \\
$\log g$ ($\rm{cm}\,\rm{s}^{-2}$) & 4.35$\pm$0.05 & \protect\cite{gray}\\
$\rm{L} (L_\odot)$ & 4.92$\pm$0.41 & \\
\hline
$\rm{v}\sin i$ (km $\rm{s}^{-1}$) & 37.5$\pm$2 & \protect\cite{kaye2}\\
\hline
$\pi$ (mas) & 25.38$\pm$0.85 & \protect\cite{leeuwen}\\
\hline
\end{tabular}
\end{table}

\cite{gray} obtained spectroscopic observations of this star,
providing accurate values for the stellar luminosity, $T_{\rm{eff}}$,
and $\log\,g$ (see Table 2). The use of different bolometric
corrections changes significantly the value of the luminosity of the
star. We have used the Virtual observatory tool VOSA \citep{amelia} to
avoide the necessity for estimating the stellar luminosity using
bolometric corrections. Models with realistic metallicities best
fitting the observations provide a luminosity only 0.1 $L/L_{\odot}$
larger than that given in \cite{gray} and, therefore, within the
errors. On the other hand, the use of different parallaxes in the
literature changes the value of the absolute magnitude of this star by
less than 0.1 $L/L_{\odot}$ (see Paper I for details).

We have developed a grid of equilibrium models obtained with the CESAM
code \citep{Morel08}, with physics appropiate for main sequence A type
stars (see Paper I). In particular, the abundance mixture used is that
given in \cite{Grevessenoels93}. To obtain the initial hydrogen and
helium abundances we have used a primordial helium content $Y_{\rm
  P}=0.235$ and a mean enrichment law $\Delta Y/\Delta Z=2.2$. The use
of other determinations of these parameter in the literature
\citep[for example]{claretwill} does not change significantly the
initial abundances used. The main approximation taken in the models
with a possible influence in this study is the absence of updated
internal chemical transport mechanisms in the equilibrium models.  We
vary the mass (in the range [1.25, 2.10] $M_\odot$ with steps of 0.01
$M_\odot$), the metallicity (with values [M/H]=0.08, $-$0.12, $-$0.32,
and $-$0.52), the Mixing-Length parameter MLT (values 0.5, 1, and
1.5), and the overshooting (values 0.1, 0.2, and 0.3). The internal
metallicity has been regarded as a free parameter due to the $\lambda$
Bootis nature of the star, which does not represent its internal
abundances. The mass, estimated to be $1.47\pm0.3$ $M_\odot$ by
\cite{gray}, has been also regarded as a free parameter since it has
not been directly determined.

If we search for models fulfilling the temperature, $\log\,g$,
luminosity and radius observed for the star, only a small amount of
them remain, all in an age range [10, 2337] Myr (Fig. 1). A 18.1$\%$
of the models has the age estimated in the literature for this star
([30, 160] Myr). The gaps in the figure are due to the mesh of our
grid of models. This result shows that the HR diagram position alone
is not an accurate method to estimate the age of HR\,8799. The reason
is the unknown internal metallicity of the star due to its $\lambda$
Bootis nature. Therefore, we need additional information to estimate
the age of this planetary system.

\begin{figure}
\includegraphics[width=84mm]{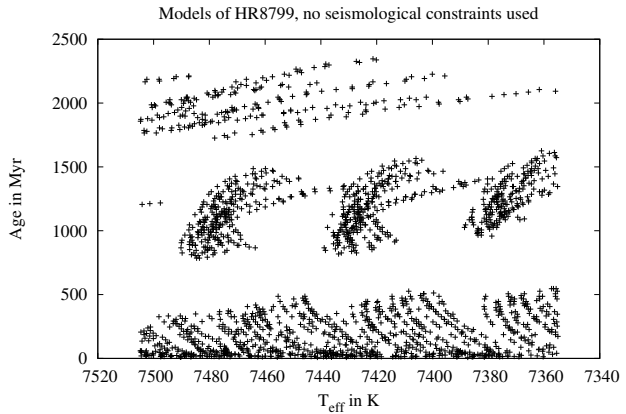}
 \caption{$T_{\rm{eff}}$ - Age diagram for the models of our grid
   fulfilling the spectroscopic observations, i.e., $T_{eff}$, $\log
   g$ and luminosity (see Table 2). The age range obtained is [10,
     2337] Myr.}
\end{figure}

A comprehensive asteroseismic modelling of HR\,8799 have been also
developed. We use the presence of $\gamma$ Doradus pulsations found in
this star. As $\gamma$ Doradus stars are asymptotic g-mode pulsators,
they are very good candidates for testing the internal structure of
the star, in particular its internal metallicity and age, both having
a large influence in the bouyancy restoring force, measured through
the Brunt- V\"ais\"al\"a frequency \citep{frm,smeyers,miglio}. To do
so, some of the most updated tools adapted for this purpose are used:
1) the evolutionary code CESAM \citep{Morel08}, and 2) the pulsation
code GraCo \citep{graco1,graco2}. The latter has been tested in the
work of the ESTA activity \citep[Evolution and Asteroseismic Tools
  Activities]{estafreq,estades}. Using these tools, we performed a
theoretical study of HR\,8799 in an attempt at constraining physical
and theoretical parameters. In this work we will follow the same
scheme used for the study of the $\delta$ Scuti pulsators RV Arietis,
29 Cygnis and 9 Aurigae \citep{Casas06,Casas09,Moya06au}. 








A description of the tools and procedures used can be found in Paper
I. In this work we focus only on the determination of the mass and age
of the star. These two quantities are the physical properties of the
star with a larger impact in the understanding of the nature of the
planetary system.

\begin{table*}
 \centering
 \begin{minipage}{160mm}
  \caption{Acceptable models depending on the physical
    constraints. In the less favorable case, the first (second)
      mass range is linked with the first (second) age range shown.}
  \begin{tabular}{cccc}
  \hline Constraint & Mass & Age & $\%$ of models with $[30, 160]$ \\
 &  in $M_{\odot}$ & in Myr & in Myr \\
\hline 
HR position & [1.25,1.27],[1.32,1.35],[1.40,1.48] & [10, 2337] & 18.1\\ 
Complete procedure & 1.32 & [1126, 1486] & 0.0 \\
Comp. proc. (less favorable case) & [1.32, 1.33], [1.44,
  1.45] & $[1123, 1625]$, $[26, 430]$ & 16.7\\ 
\hline
\end{tabular}
\end{minipage}
\end{table*}

An important unknown for the present study is the rotation
velocity. \cite{kaye2} gave a projected rotation velocity for this
star of $v\sin i=37.5\pm 2.0\,\rm{km}\,\rm{s}^{-1}$. Therefore, an
accurate determination of $v$ moves to an accurate determination of
the visual angle $i$. Several works in the literature tried to provide
estimations for $i$ with no conclusive results. This is an important
condition for the modelling.

In the present study, we use the Frequency Ratio Method (FRM)
developed in \cite{frm} (see Section 3.3 of Paper I) for the
asteroseismological modelling of this star. The FRM is based on the
analytical description of the frequencies in the asymptotic
regime. This procedure has a limited range of validity in stellar
rotation velocities, as demonstrated in \cite{frmrot}. In that paper,
the authors found a maximum rotation velocity for the correct
application of the FRM, for a standard $\gamma$ Doradus star, of
around $60\,\rm{km}\,\rm{s}^{-1}$. On the other hand, \cite{turcotte}
found that meridional circulation cannot destroy the accretion pattern
(in the accretion/diffusion scenario, based on the accretion of
inter-stellar medium by the star, and the mixing of these elements in
the stellar surface due to diffusion and rotationally mixing
processes) for a rotation velocity below
$125\,\rm{km}\,\rm{s}^{-1}$. For a larger rotation velocity, theory
cannot ensure that the accreted abundances can remain in the stellar
surface enough time to be observed. These values provide two lower
limits for the inclination angle.

Obtaining the rotation velocity as a function of $i$ for any stellar
radius in the observed range we found that $i>18^\circ$ would be the
requirement imposed by the $\lambda$ Bootis nature of HR\,8799, and
$i>36^\circ$ would be the requirement for the FRM to be applicable. We
want to point out that if the inclination angle is between both
values, our analysis would be possible but inaccurate, since we
  cannot ensure that the real solution is part of that provided by the
  FRM \citep{frmrot}.

\begin{figure*}
\includegraphics[width=84mm]{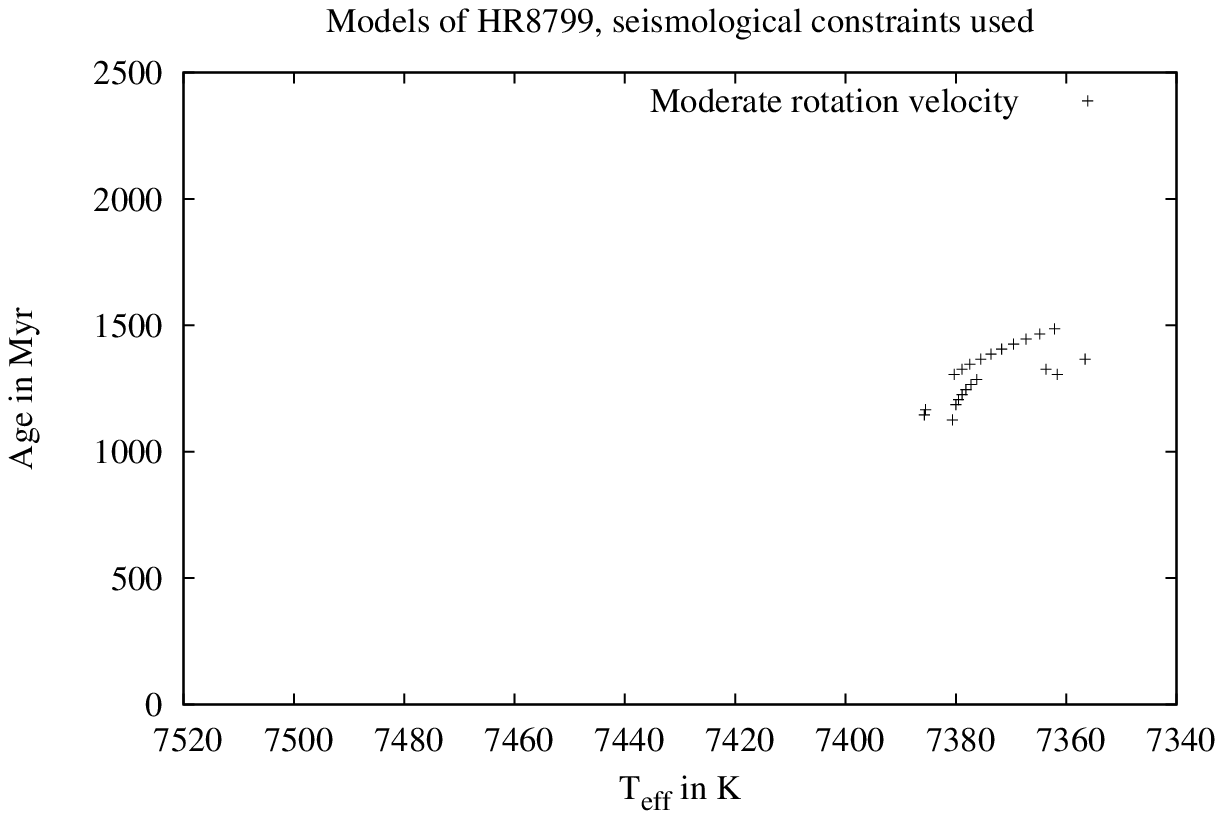}
\includegraphics[width=84mm]{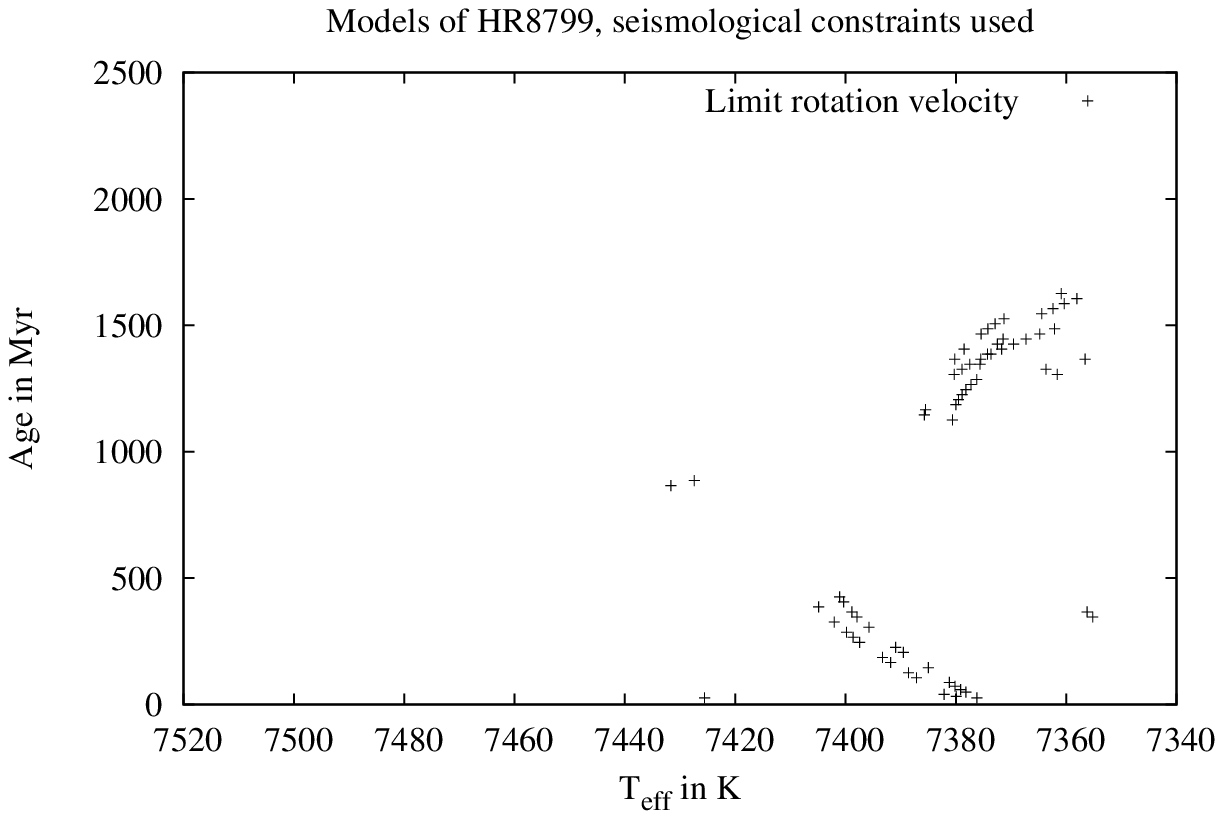}
 \caption{$T_{\rm{eff}}$ - Age diagram for the models of our grid
   fulfilling the spectroscopic observations plus all the
   asteroseismological constraints (FRM + multicolour photometry
     + instability analysis, see Paper I) for a rotation velocity
   $V_{\rm{rot}}\approx 45 \rm{km}\,\rm{s}^{-1}$ (left panel) and
   $V_{\rm{rot}}\approx 60 \rm{km}\,\rm{s}^{-1}$ (right panel).
     The age ranges obtained are $[1123, 1625]$ Myr and
     $[1123, 1625]$, $[26, 430]$ Myr, respectively}
\end{figure*}

Our first study was done for a moderate rotation case in the Frequency
Ratio Method. This means an inclination angle $i\approx 50^\circ$ and
a rotation velocity $V_{\rm{rot}}\approx 45\,\rm{km}\,\rm{s}^{-1}$. In
this case, the complete procedure is very discriminant and only models
with M$=1.32\,M_{\odot}$ fulfill all the observational
constraints. The acceptable age range is [1126, 1486] Myr, far from
the value estimated for this star by \cite{Marois} ([30,160] Myr). In
Fig. 2 (left panel), the acceptable set of models for this rotation
velocity is shown in a $T_{\rm{eff}}$-Age diagram. The $T_{\rm{eff}}$
axis limits are provided by the spectroscopic observations.

We have also studied the less favourable case in the FRM, i.e. an
inclination angle $i\approx 36^\circ$, that is, $V_{\rm{rot}}\approx
60\,\rm{km}\,\rm{s}^{-1}$. In this case, the complete procedure is not
completely discriminant and new models fulfill all the observational
constraints. Two mass ranges are now acceptable (M=[1.32, 1.33] and
[1.44, 1.45] $M_\odot$), and the compatible age range is divided into
two blocks: [1123, 1625] Myr and [26, 430] Myr respectively
(Fig. 2, right panel). The two models with ages around 800 Myr have
been accepted due to the conservative application of the instability
analysis (see Paper I), but it their belonging to the set of
acceptable models in unlikely. In this less favourable case, the
percentage of accepted models with ages in the range commonly used for
this star is 16.7$\%$. A summary of the results presented here is found
in Table 3.

\section{Conclusions and future perspectives}

In the present work, an analysis of the age determination of the
planetary system HR\,8799 has been done. The results found in the
literature are not conclusive, and the only valid argument to
estimate the age of the star is that using its radial velocity and
proper motion, but it is an estatistical argument needed of additional
estimations.

The only valid argument used to estimate the age of the star is its
radial velocity and proper motion, but it is a statistical argument
needed of additional estimations. The main complementary argument is
the position of the star in the HR diagram. In this work we have
demostrated that this procedure does not provide accurate age
estimations due to the $\lambda$ Bootis nature of HR\,8799. This
nature hides the real internal metallicity of the star. The main
consequence of this result is that the models fulfilling observations
are in a range of ages [10,2337] Myr, a much broader range that
one estimated by other authors of [30,160] Myr \citep{Marois}. Only a
small amount (18.1$\%$) of models in our representative grid have ages
in the range claimed in the literature.

Therefore we need aditional constraints for an accurate estimation of
the age and mass of the star (these two quantities are the physical
characteristics of the star with larger impact in the understanding of
the planetary system). We have taken advantage of the $\gamma$
Doradus pulsations of the star to better estimate these values with
the help of asteroseismology.

A comprehensive asteroseismological study of this star has been
developed. This study is described in detail in \cite[Paper
  I]{hr8799lambda}. The main source of uncertainty of the procedure is
the unknown rotation velocity of the star. We have analysed the
possible results depending on the inclination angle $i$. There is a
range of angles where this study is not accurate
($i=[18^\circ,36^\circ]$).

For angles around $i=36^\circ$, the models fulfilling all the
observational constraints have masses in two separate ranges of
M=[1.32, 1.33], [1.44, 1.45] $M_{\odot}$. The age of the system is
constrained in two separate ranges: [1123, 1625] Myr and [26, 430] Myr
respectively. A percentage of 16.7$\%$ of the models are in the range
given in the literature, i.e., young ages. A consequence of this
result is that, in the case of the youngest age range, the predicted
masses of the observed planets are [5,14] $M_{\rm{Jup}}$ for the most
luminous planets, and [3,13] $M_{\rm{Jup}}$ for the less luminous
one. The oldest age range predicts masses for the three objects in the
brown dwarfs domain \citep[see Fig. 4 of][]{Marois}.

In the most favourable case for the procedure used in this work,
i.e. inclination angles of around $i\approx 50^\circ$,
asteroseismology is very accurate, and the star would have an age in
the range [1126, 1486] Myr, and a mass M=1.32 $M_{\odot}$.  This age
range implies that the observed objects orbiting HR\,8799 would be
brown dwarfs \citep[following Fig. 4 in][]{Marois}.

The lack of an accurate determination of the inclination angle is the
main source of uncertainty of the present study.  This angle has not
been unambiguously obtained up to now, and its value would say whether
the results of this study are actually applicable, and then, it can
provide a very accurate determination of the age and mass of the
star. In any case, one of the main conclusions of our study is that
the range of ages assigned to this star in the literature is unlikely
to be the correct one. Only a stellar luminosity larger than that
reported would allow young models with solar metallicity to fulfill
all the observational constraints.

\section*{Acknowledgements}

AM acknowledges financial support from a Juan de la Cierva contract of
the Spanish Ministry of Science and Innovation. PJA acknowledges
financial support from a ``Ramon y Cajal'' contract of the Spanish
Ministry of Education and Science. This research has been funded by
Spanish grants ESP2007-65475-57-C02-02 , CSD2006-00070,
ESP2007-65480-C02-01, AYA2009-08481-E and CAM/PRICIT-S2009ESP-1496.

\end{document}